\documentclass[aps,pra,twocolumn,amssymb,amsmath]{revtex4-1}
\usepackage{graphicx} 
\usepackage{bm}

\begin{document}
\title{Traveling waves of nonlinear Schr\"{o}dinger equation including higher order dispersions}

\author{Vladimir I. Kruglov}

\affiliation{Centre for Engineering Quantum Systems,
School of Mathematics and Physics, The University of Queensland, Brisbane, QLD 4072, Australia.}

\begin{abstract}
The solitary wave solution and periodic solutions expressed in terms of elliptic Jacobi's functions are obtained for the nonlinear Schr\"{o}dinger equation governing the propagation of pulses in optical fibers including the effects of second, third and fourth order dispersion. The approach is based on the reduction of the generalized nonlinear Schr\"{o}dinger equation to an ordinary nonlinear differential equation. The periodic solutions obtained form one-parameter family which depend on an integration constant $p$. The solitary wave solution with ${\rm sech}^2$ shape is the limiting case of this family with $p=0$. The solutions obtained describe also a train of soliton-like pulses with ${\rm sech}^2$ shape. It is shown that the bounded solutions
arise only for special domains of integration constant. 
\end{abstract}

\maketitle

\section{Introduction}

The nonlinear Schr\"{o}dinger equation governs the dispersive propagation of a pulse envelope with a high carrier frequency in a weakly nonlinear medium. For example,
the propagation of short intense optical pulses in nonlinear dispersive media leads to the formation of optical solitons \cite{HT}.
The nonlinear Schr\"{o}dinger equation (NLSE) has been used to
model the dynamics of a Bose-Einstein condensate \cite{G1,G2,G3}, dispersive 
shock waves in hydrodynamic media \cite{G4} and EM propagation
in optical waveguides \cite{AG}. 
The application of NLSE to light propagation in optical fibres has led to an understanding of a rich variety of phenomena, including optical wave breaking, modulation instability and the generation of solitary waves 
and parabolic pulses (similaritons) \cite{G7,KR,KH}.
Whilst the limitation to a consideration of second order dispersion only is adequate for many situations,
the advent of photonic crystal waveguides and silicon photonic structures with complex dispersion characteristics has led to more realistic description of the dispersion of the waveguide with the higher order terms in the expansion of the propagation constant.
Solitary waves governed by second and fourth order dispersion only (quartic solitons), have been studied \cite{B1,B2,B3,B4,B5}. Moreover, it is shown by numerical simulsations that the temporal shape and the peak power of the soliton are stable when a weak third order dispersion is introduced \cite{MJ}. The exact solutions are also obtained for high dispersive NLSE in the next papers     \cite{B6,MT,GS,ZH}.

The recent advent of silicon photonics has provided a wide range of dispersion profiles wherein the propagation of pulses is described by the NLSE \cite{C1,C2,C3,C4,C5}.
Experimental and numerical evidence for pure-quartic solitons and periodically modulated propagation for the higher-order quartic soliton has also been reported \cite{D1,AB}
while optical pulses in photonic crystal wavegides have also been observed exhibiting slow light propagation at particular frequencies \cite{G8}.
An exact stationary soliton-like solution of the generalized NLSE with second, third and fourth order dispersion terms has recently been obtained,  together with an approximate solution in the case when an absorption or gain term is included in the equation \cite{KRH}.

In this paper we present the approach which yields one-parameter family of bounded solutions of the generalized nonlinear Schr\"{o}dinger equation with second, third and fourth order dispersion terms. This family of solutions depend on some integration constant $p$ and is expressed in terms of elliptic Jacobi's functions. The domain of
integration constant $p$ for bounded solutions depends on the sign of nonlinear parameter $\gamma$ in the generalized NLSE. In nonlinear fiber optics the parameter $\gamma$ is positive. 
However we consider in this paper also the case with negative parameter $\gamma$ (defocusing nonlinearity). Note that the class of bounded solutions may find application in other fields of physics where the generalized NLSE is a good model of a system.
Here the family of exact solutions is referred to as ${\cal P}_{+}$ class of bounded solutions. The solitary wave solution with ${\rm sech}^2$ shape is the limiting case of this family when the integration constant $p$ is zero and the parameters in the generalized NLSE satisfy some necessary conditions. We refer below, for simplicity, to the solitary wave as a ${\rm sech}^2$  soliton. In the case when some dimensionless  
parameter connected to the integration constant $p$ is small, the solutions obtained describe a train of soliton-like optical pulses with ${\rm sech}^2$ shape. In the case when the integration constant $p$ is zero the train of soliton-like optical pulses reduces to the solitary wave solution with ${\rm sech}^2$ shape. It is shown that the generalized NLSE does not have a quartic dark soliton solution. Moreover we show in this paper that the traveling bounded solutions
of the generalized NLSE arise for integration constant $p\geq 0$ and $p\leq 0$ when $\gamma>0$ and $\gamma<0$ respectively. Five Statements are proved in this paper which asserts some basic propositions.

\section{Traveling waves of nonlinear Schr\"{o}dinger equation}

In this section we present the method which leads to the reduction of the generalized NLSE to an ordinary differential equation whose solutions can be expressed in terms of elliptic Jacobi's functions. The generalized nonlinear Schr\"{o}dinger equation is derived for the assumptions of slowly varying envelope, instantaneous nonlinear response, and no higher order nonlinearities \cite{A1,A2,A3}. 
This nonlinear equation has the next form for the optical pulse envelope $\psi(z,\tau)$,
\begin{equation}
i\frac{\partial\psi}{\partial z}=\alpha\frac{\partial^2\psi}{\partial \tau^2}+i\sigma\frac{\partial^3\psi}{\partial \tau^3} 
-\epsilon\frac{\partial^4\psi}{\partial \tau^4}-\gamma |\psi|^2\psi,
\label{1}
\end{equation}
where $z$ is the longitudinal coordinate, $\tau=t-\beta_1z$ is the retarded time, and 
$\alpha=\beta_2/2$, $\sigma=\beta_3/6$, $\epsilon=\beta_4/24$, and $\gamma$ is the nonlinear parameter. The parameters 
$\beta_k=(d^k\beta/d\omega^k)_{\omega=\omega_0}$ are the k-order dispersion of the optical fiber and $\beta(\omega)$ is the propagation constant depending on the optical frequency.
We consider the solution of the generalized NLSE in the form,
\begin{equation}
\psi(z,\tau)={\rm u}({\rm x})\exp[i(\kappa z-\delta\tau+\phi)],
\label{2}
\end{equation}
where ${\rm u}({\rm x})$ is a real function depending on the variable ${\rm x}=\tau-qz$, and $q={\rm v}^{-1}$ is the inverse velocity. Equations (\ref{1}) and (\ref{2}) lead to the next system of the ordinary differential equations,
\begin{equation}
(\sigma+4\epsilon\delta)\frac{{\rm d}^3{\rm u}}{{\rm d}{\rm x}^3}+(q-2\alpha\delta-3\sigma\delta^2-4\epsilon\delta^3)\frac{{\rm d}{\rm u}}{{\rm d}{\rm x}}=0,
\label{3}
\end{equation}
\begin{eqnarray}
\epsilon\frac{{\rm d}^4{\rm u}}{{\rm d}{\rm x}^4}-(\alpha+3\sigma\delta+6\epsilon\delta^2)\frac{{\rm d}^2{\rm u}}{{\rm d}{\rm x}^2}+\gamma{\rm u}^3~~~~~~~~~~
\nonumber\\ \noalign{\vskip3pt}  -(\kappa-\alpha\delta^2-\sigma\delta^3-\epsilon\delta^4){\rm u}=0.~~~~~~~~~~~~~~~~
\label{4}
\end{eqnarray}
In the general case the system of Eqs. (\ref{3}) and (\ref{4}) is overdetermined because we have two differential equations for the function ${\rm u}({\rm x})$.
However, if some constraints for the parameters in Eq. (\ref{3}) are fulfilled the system of Eqs. (\ref{3}) and (\ref{4}) has non-trivial solutions. We refer the solution of the generalized NLSE where the function $\psi(z,\tau)$ is given by Eq. (\ref{2}) with ${\rm u}({\rm x})\neq {\rm constant}$ as non-plain wave or traveling wave solution. 

{\it Statement 1}. The system of Eqs. (\ref{3}) and (\ref{4}) with $\epsilon\neq 0$ yields the non-plain wave solutions if and only if the next relations are satisfied: 
\begin{equation}
q=2\alpha\delta+3\sigma\delta^2+4\epsilon\delta^3,~~~~\delta=-\frac{\sigma}{4\epsilon}.
\label{5}
\end{equation}
The system of Eqs. (\ref{3}) and (\ref{4}) with $\epsilon=0$ has the non-plain wave solutions only when the parameter $\sigma=0$.

The proof of this Statement is given in Appendix A. Note that Eq. (\ref{3}) is satisfied for an arbitrary function ${\rm u}({\rm x})$ according to the conditions in Eq. (\ref{5}) with $\epsilon\neq 0$.
The relations in Eq. (\ref{5}) lead to the next expression for the velocity ${\rm v}=1/q$ defined in the retarded frame,  
\begin{equation}
{\rm v}=\frac{8\epsilon^2}{\sigma(\sigma^2-4\alpha\epsilon)}.
\label{6}
\end{equation}
The relations in Eq. (\ref{5}) reduce the system of Eqs. (\ref{3}) and (\ref{4}) to
the ordinary nonlinear differential equation, 
\begin{equation}
\epsilon\frac{{\rm d}^4{\rm u}}{{\rm d}{\rm x}^4}+b\frac{{\rm d}^2{\rm u}}{{\rm d}{\rm x}^2}-c{\rm u}+\gamma{\rm u}^3=0,~~~~~~~
\label{7}
\end{equation}
where the parameters $b$ and $c$ are
\begin{equation}
b=\frac{3\sigma^2}{8\epsilon}-\alpha,~~~~c=\kappa+\frac{\sigma^2}{16\epsilon^2}\left(
\frac{3\sigma^2}{16\epsilon}-\alpha\right).
\label{8}
\end{equation}
The multiplication of Eq. (\ref{7}) by the function ${\rm d}{\rm u}/{\rm d}{\rm x}$ and integration of this equation leads to the next third order nonlinear differential equation,
\begin{equation}
\frac{{\rm d}{\rm u}}{{\rm d}{\rm x}}\frac{{\rm d}^3{\rm u}}{{\rm d}{\rm x}^3}
-\frac{1}{2}\left(\frac{{\rm d}^2{\rm u}}{{\rm d}{\rm x}^2}\right)^2+
\frac{b}{2\epsilon}\left(\frac{{\rm d}{\rm u}}{{\rm d}{\rm x}}\right)^2
-\frac{c}{2\epsilon}{\rm u}^2+\frac{\gamma}{4\epsilon}{\rm u}^4=\frac{\Omega}{\epsilon},
\label{9}
\end{equation}
where $\Omega$ is an integration constant. We introduce a new function ${\rm F}({\rm u})$
defined by equation,
\begin{equation}
\left(\frac{{\rm d}{\rm u}}{{\rm d}{\rm x}}\right)^2={\rm F}({\rm u}).
\label{10}
\end{equation}
The system of Eqs. (\ref{9}) and (\ref{10}) yields the next second order nonlinear differential equation,
\begin{equation}
{\rm F}\frac{{\rm d}^2{\rm F}}{{\rm d}{\rm u}^2}
-\frac{1}{4}\left(\frac{{\rm d}{\rm F}}{{\rm d}{\rm u}}\right)^2+\nu{\rm F}
+\eta_0+\eta_1{\rm u}^2+\eta_2{\rm u}^4=0,
\label{11}
\end{equation}
where $\nu=b/\epsilon$, $\eta_0=-2\Omega/\epsilon$, $\eta_1=-c/\epsilon$ and $\eta_2=\gamma/2\epsilon$.
It is shown in Appendix B that Eq. (\ref{11}) has the polynomial solution as
\begin{equation}
{\rm F}({\rm u})=a_0+a_1{\rm u}+a_2{\rm u}^2+a_3{\rm u}^3,
\label{12}
\end{equation}
where the coefficients $a_n$ and integration constant $\Omega$ are
\begin{equation}
a_0=\frac{2bp}{9\gamma\epsilon},~~~~a_1=\pm\frac{2p}{9\epsilon}\sqrt{-\frac{15\epsilon}{2\gamma}},
~~~~a_2=-\frac{b}{5\epsilon},
\label{13}
\end{equation}
\begin{equation}
a_3=\pm\sqrt{-\frac{2\gamma}{15\epsilon}},~~~~\Omega=\frac{3b^2p}{45\gamma\epsilon}+\frac{5p^2}{108\gamma}.
\label{14}
\end{equation}
Here $p$ is an arbitrary real parameter connected to wave number as $\kappa=p+\kappa_0$ (see Appendix B) where the constant parameter $\kappa_0$ is
\begin{equation}
\kappa_0=-\frac{4}{25\epsilon^3}\left(\frac{3\sigma^2}{8}-\alpha\epsilon \right)^2-\frac{\sigma^2}{16\epsilon^3}\left(\frac{3\sigma^2}{16}-\alpha\epsilon \right).
\label{15}
\end{equation}
Thus the wave number $\kappa$ in Eq. (\ref{2}) is given by
\begin{equation}
\kappa=p-\frac{4}{25\epsilon^3}\left(\frac{3\sigma^2}{8}-\alpha\epsilon \right)^2-\frac{\sigma^2}{16\epsilon^3}\left(\frac{3\sigma^2}{16}-\alpha\epsilon \right).
\label{16}
\end{equation}
We emphasise that the wave number $\kappa$ is the integration constant in Eq. (\ref{2}) and hence the parameter $p$ is a new integration constant defined as $p=\kappa-\kappa_0$. 
Thus the system of Eqs. (\ref{10}) and (\ref{12}) leads to the next ordinary nonlinear differential equation,
\begin{equation}
\left(\frac{{\rm d}{\rm u}}{{\rm d}{\rm x}}\right)^2=a_0+a_1{\rm u}+a_2{\rm u}^2+a_3{\rm u}^3,
\label{17}
\end{equation}
where the coefficients $a_n$ and the integration constant $\Omega$ are given in Eqs. (\ref{13}) and (\ref{14}) with $p=\kappa-\kappa_0$. Moreover the constant parameter $\kappa_0$ and wave number $\kappa$ are defined by Eqs. (\ref{15}) and (\ref{16}).

In nonlinear fiber optics the nonlinear parameter is positive ($\gamma>0$). However Eq. (\ref{1}) may find also application in other fields of physics where the nonlinear parameter is negative ($\gamma<0$). Thus we consider below these two cases. It follows from Eqs. (\ref{13}) and (\ref{14}) that in the first case
the nonlinear parameter $\gamma>0$ and hence $\epsilon<0$, and in the second case the parameter 
$\gamma<0$ and hence $\epsilon>0$. We consider for generality the non-plain wave solutions of Eq. (\ref{1}) for condition ${\rm C}_1$ ($\gamma>0$, $\epsilon<0$) and condition ${\rm C}_2$ ($\gamma<0$, $\epsilon>0$). 

{\it Statement 2}. All non-plain wave solutions  of the first order ordinary nonlinear differential equation (\ref{17}) are the solutions of Eq. (\ref{7}) provided that the coefficients $a_n$, integration constant $\Omega$, parameter $\kappa_0$ and wave number $\kappa$ are given by Eqs. (\ref{13})-(\ref{16}).

We have shown that Eq. (\ref{17}) follows from Eq. (\ref{7}) provided that the integration constant $\Omega$,  parameter $\kappa_0$ and wave number $\kappa=p+\kappa_0$, where $p$ is an arbitrary real parameter, are given by Eqs. (\ref{13})-(\ref{16}). 
 Let the function ${\rm u}({\rm x})$ is differentiable to the fourth order inclusively then
the second and fourth order derivatives of the function ${\rm u}({\rm x})$ follow from Eq. (\ref{17}). 
 The substitution of these derivatives to Eq. (\ref{7}) provided that ${\rm u}({\rm x})\neq {\rm constant}$ leads to coefficients $a_n$ given in Eqs. (\ref{13}) and (\ref{14}) where $p=\kappa-\kappa_0$. Moreover the constant parameter $\kappa_0$ and wave number $\kappa$ are given by Eqs. (\ref{15}) and (\ref{16}). This complete the proof of this Statement.
  
The integration of Eq. (\ref{17}) yields the general solution, 
\begin{equation}
\pm\int \frac{d{\rm u}}{\sqrt{a_0+a_1{\rm u}+a_2{\rm u}^2+a_3{\rm u}^3}}={\rm x}+C,          
\label{18}
\end{equation}
where $C$ is an integration constant.

\section{Solitary wave and plain wave solutions}

In this section we derive the solitary wave solution of the generalized NLSE with ${\rm sech}^2$ shape \cite{KRH}. Moreover it is shown that the generalized NLSE does not have a quartic dark solution solution. The plain wave solution is also presented in this section.

It follows from Eq. (\ref{17}) that the solitary wave solution 
${\rm u}({\rm x})=A_0~{\rm sech}^2[{\rm w}_0({\rm x}-\eta)]$ takes place only in the case when $a_0=0$, $a_1=0$ and $a_2>0$. Hence the solitary wave solution exists only in the case when the integration constant is $p=0$ and $a_2>0$. The condition $a_2>0$ yields the relation $8\alpha\epsilon>3\sigma^2$, and hence the localized soliton-like solution of Eq. (\ref{17}) for $p=0$ and $a_2>0$ (see also Eqs. (\ref{11b}) and (\ref{12b})) is given by
\begin{equation}
{\rm u}({\rm x})=-\left(\frac{a_2}{a_3}\right)~{\rm sech}^2\left(\frac{\sqrt{a_2}}{2}({\rm x}-\eta)\right).
\label{19}
\end{equation}
This solitary wave solution we call below as ${\rm sech}^2$ soliton. 
Note that Eq. (\ref{17}) has also the unbounded solution as ${\rm u}({\rm x})=-(a_2/a_3)~\sec^2[\tilde{{\rm w}}_0({\rm x}-\eta)]$ where $\tilde{{\rm w}}_0=
\sqrt{|a_2|}/2$ and $a_2<0$. This solution follows from Eq. (\ref{19}) in the case when $a_2<0$ (see also Eq. (\ref{13b})). However, only the bounded solutions   
have physical applications. 
  
The solution in Eq. (\ref{19}) and the ansatz given by Eq. (\ref{2}) for $p=0$ lead to ${\rm sech}^2$ soliton solution \cite{KRH} as 
\begin{equation}
\psi(z,\tau)={\rm U}_0~{\rm sech}^2({\rm w}_0\xi)\exp[i(\kappa_0 z-\delta\tau+\phi)],
\label{20}
\end{equation}
where $\xi=\tau-{\rm v}^{-1}z-\eta$. The constant parameters $\eta$ and $\phi$ represent the position and phase of the localized pulse at $z=0$.
The amplitude and inverse temporal width of the solitary wave are given by
\begin{equation}
{\rm U}_0=\sqrt{\frac{-3}{10\gamma\epsilon}}\left(\frac{3\sigma^2}{8\epsilon}-\alpha \right),~~~
{\rm w}_0=\frac{1}{4}
\sqrt{\frac{4\alpha}{5\epsilon}-\frac{3\sigma^2}{10\epsilon^2}}.
\label{21}
\end{equation}
The velocity ${\rm v}$ of the solitary wave in the retarded frame and the parameters $\delta$ and $\kappa_0$ are
\begin{eqnarray}
{\rm v}=\frac{8\epsilon^2}{\sigma(\sigma^2-4\alpha\epsilon)},~~~~~~~~~~~~\delta=-\frac{\sigma}{4\epsilon},
\nonumber\\ \noalign{\vskip3pt} \kappa_0=-\frac{4}{25\epsilon^3}\left(\frac{3\sigma^2}{8}-\alpha\epsilon \right)^2-\frac{\sigma^2}{16\epsilon^3}\left(\frac{3\sigma^2}{16}-\alpha\epsilon \right).
\label{22}
\end{eqnarray}
The wave number $\kappa_0$ is found by Eq. (\ref{16}) with $p=0$. 
Note that $\delta$ and $\kappa_0+\beta_1\delta$ are the frequency and wave number shifts respectively.

It follows from Eq. (\ref{21}) that  the necessary and sufficient conditions for existence of ${\rm sech}^2$ solitary wave are given by two conditions as (1) $\gamma\epsilon<0$, (2)
$8\alpha\epsilon>3\sigma^2$ \cite{NS}. Hence these necessary and sufficient conditions yield
$\alpha<0$, $\epsilon<0$ and $8\alpha\epsilon>3\sigma^2$ for $\gamma>0$. These relations can also be written as $\beta_2<0$, $\beta_4<0$, and $2\beta_2\beta_4>\beta_3^2$ for $\gamma>0$, where $\beta_3$ is  negative, positive or zero. In addition these necessary and sufficient conditions yield $\alpha>0$, $\epsilon>0$ and $8\alpha\epsilon>3\sigma^2$ for $\gamma<0$ (defocusing nonlinearity).

One can also assume the existence of bounded solution of Eq. (\ref{17}) in the next form,
\begin{equation}
{\rm u}({\rm x})=A+B~{\rm th}^2[{\rm w}({\rm x}-\eta)].
\label{23}
\end{equation}
We refer below to the function ${\rm u}({\rm x})$ of this form with $A\neq -B$ as a quartic dark soliton. Note that in the case $A=-B$ this function reduces to ${\rm sech}^2$ soliton. It is straightforward to show that the function in Eq. (\ref{23}) is the solution of
Eq. (\ref{17}) only in the case when the next relations are satisfied, 
\begin{eqnarray}
\nu_0=-AG^2,~~~~\nu_1=G^2+2AG,
\nonumber\\ \noalign{\vskip3pt} \nu_2=-(A+2G),~~~~a_3B=4{\rm w}^2,
\label{24}
\end{eqnarray}
where $G=A+B$ and $\nu_n=a_n/a_3$ for $n=0, 1, 2$. These four equations yield four parameters: A, B, ${\rm w}$, and the integration constant $p$. The solution of these equations leads to the next relations,  
\begin{equation}
G=-\frac{\nu_2}{3}\pm\frac{1}{3}\sqrt{\nu_2^2-3\nu_1},~~~~{\rm w}=\frac{1}{2}\sqrt{a_3B},
\label{25}
\end{equation}
\begin{equation}
A=-\frac{\nu_2}{3}\mp\frac{2}{3}\sqrt{\nu_2^2-3\nu_1},~~~~B=\pm\sqrt{\nu_2^2-3\nu_1},
\label{26}
\end{equation}
where $\nu_2^2-3\nu_1>0$.
The relations in Eq. (\ref{24}) yield the equation for integration constant $p$
as
\begin{equation}
9\nu_0-3\nu_1\nu_2+\frac{2}{3}\nu_2^3=\pm\frac{2}{3}(\nu_2^2-3\nu_1)^{3/2},
\label{27}
\end{equation}
where the coefficients $\nu_n=a_n/a_3$ are
\begin{equation}
\nu_0=\pm\frac{2bp}{9\gamma\epsilon}\sqrt{-\frac{15\epsilon}{2\gamma}},~~~\nu_1=-\frac{5p}{3\gamma},~~~\nu_2=\mp\frac{b}{5\epsilon}\sqrt{-\frac{15\epsilon}{2\gamma}},
\label{28}
\end{equation}
with $\gamma\epsilon<0$.
Thus Eq. (\ref{27}) has the next explicit form,
\begin{equation}
\sqrt{-\frac{15\epsilon}{2\gamma}}\left(\frac{3b^3}{50\gamma\epsilon^2}+\frac{3bp}{2\gamma\epsilon}\right)=\left(-\frac{3b^2}{10\gamma\epsilon}+\frac{5p}{\gamma}\right)^{3/2}.
\label{29}
\end{equation}
We consider below the parameter $b\neq 0$ because this equation yields $p=0$ and ${\rm w}=0$ for $b=0$.
In the case when $b\neq 0$ Eq. (\ref{29}) has the dimensionless form,
\begin{equation}
-\frac{\epsilon b}{|\epsilon b|}(1-P)=\left(1+\frac{2}{3}P\right)^{3/2},~~~~P=-\frac{25\epsilon}{b^2}p.
\label{30}
\end{equation}
This equation has a real solution $P=0$ when $\epsilon b<0$. Thus $p=0$ is a solution of 
Eq. (\ref{30}) for $\gamma\epsilon<0$ and $8\alpha\epsilon>3\sigma^2$ ($\nu_2^2-3\nu_1>0$).
The cubic equation follows from Eq. (\ref{30}) as $P^3+(9/8)P^2+(27/2)P=0$. 
This equation has one real solution $P=0$ and two complex conjugated solutions. Moreover
each solution of Eq. (\ref{30}) is also a solution of this cubic equation. However the cubic equation has only one real solution $P=0$, and hence Eq. (\ref{29}) has also only one
real solution $p=0$. Thus Eqs. (\ref{25}) and (\ref{26})
for $\gamma\epsilon<0$, $8\alpha\epsilon>3\sigma^2$ and $p=0$ yield the next parameters in Eq. (\ref{23}):
$A=-a_2/a_3$, $B=a_2/a_3$ and ${\rm w}=\sqrt{a_2}/2$. These parameters reduce the function
${\rm u}({\rm x})$ in Eq. (\ref{23}) to ${\rm sech}^2$ soliton given in Eq. (\ref{19}). Hence
Eq. (\ref{17}) does not have a quartic dark soliton solution for ${\rm C}_1$ and ${\rm C}_2$  conditions.

{\it Statement 3}. The generalized NLSE in Eq. (\ref{1}) does not have a quartic dark soliton solution. The function in Eq. (\ref{23}) is the traveling solution of the generalized NLSE only in the case when the integration constant is $p=0$ which yields
${\rm sech}^2$ soliton solution given in Eq. (\ref{20}). 

It is shown above that the function in Eq. (\ref{23}) is the solution of Eq. (\ref{17}) only in the case when the integration constant is $p=0$. In this case the function in Eq. (\ref{23}) reduces to ${\rm sech}^2$ soliton solution in Eq. (\ref{19}). Moreover 
the coefficients in Eq. (\ref{17}) have a unique form given
by Eqs. (\ref{13}) and (\ref{14}) which complete the proof of this Statement. 

The plain wave solution of the generalized NLSE follows from Eq. (\ref{4}) as
\begin{equation}
\psi(z,\tau)=\sqrt{\gamma^{-1}(\kappa-\alpha\delta^2-\sigma\delta^3-\epsilon\delta^4)}~\exp[i(\kappa z-\delta\tau+\phi)],
\label{31}
\end{equation}
where all parameters are the arbitrary real numbers satisfying the next condition 
$\gamma^{-1}(\kappa-\alpha\delta^2-\sigma\delta^3-\epsilon\delta^4)\geq 0$.

\section{${\cal P}_{+}$ class of bounded solutions of generalized NLSE}

In this section we describe the ${\cal P}_{+}$ class of bounded solutions of generalized NLSE for ${\rm C}_1$ ($\gamma>0$, $\epsilon<0$) and ${\rm C}_2$ ($\gamma<0$, $\epsilon>0$) conditions. We show that the domains of integration constant in these two cases are $p\geq 0$ and $p\leq 0$ respectively.
Note that in these both cases the domain of integration constant $p$ for the ${\cal P}_{+}$ class of bounded solutions is given by relation $\gamma p\geq0$. 

We introduce a new function ${\rm y}({\rm x})$ by the next relation ${\rm u}({\rm x})=-(1/a_3){\rm y}({\rm x})$ then Eq. (\ref{17}) has the simplified form as
\begin{equation}
\left(\frac{{\rm d}{\rm y}}{{\rm d}{\rm x}}\right)^2=f({\rm y}),~~~~       f({\rm y})=c_0+c_1{\rm y}+c_2{\rm y}^2-{\rm y}^3,
\label{32}
\end{equation}
where the coefficients $c_n$ are
\begin{equation}
c_0=-\frac{4bp}{135\epsilon^2},~~~c_1=-\frac{2p}{9\epsilon},~~~c_2=
-\frac{b}{5\epsilon}.
\label{33}
\end{equation}
The function ${\rm u}({\rm x})$ in Eq. (\ref{2}) is given by  
\begin{equation}
{\rm u}({\rm x})=\mp\sqrt{-\frac{15\epsilon}{2\gamma}}~{\rm y}({\rm x}).
\label{34}
\end{equation}
Note that an arbitrary sign in this equation leads to the same solution of the generalized NLSE because the function $\psi(z,\tau)$ in Eq. (\ref{2}) depends on an arbitrary constant phase 
$\phi$.
The general solution of Eq. (\ref{32}) is
\begin{equation}
\pm\int \frac{d{\rm y}}{\sqrt{f({\rm y})}}={\rm x}+C,~~~~f({\rm y})=\prod_{n=1}^{3}
({\rm y}_n-{\rm y}),          
\label{35}
\end{equation}
where $C$ is integration constant and ${\rm y}_n$ are the roots of equation $f({\rm y})=0$.
It is known that the character of roots ${\rm y}_n$ in this equation depends on the sign of function $D(p)$ (see Eq. (\ref{2c})). This function has the next explicit form,
\begin{equation}
D(p)=\left(\frac{bp}{135\epsilon^2}+\frac{b^3}{(15\epsilon)^3}\right)^2+
\left(\frac{2p}{27\epsilon}-\frac{b^2}{(15\epsilon)^2}\right)^3.
\label{36}
\end{equation}
We can also present this function for $b\neq 0$ as
\begin{equation}
D(p)=-\frac{8}{27}\left(\frac{b}{15\epsilon}\right)^6Q(P),
\label{37}
\end{equation}
where $P$ is the dimensionless parameter and $Q(P)$ is the cubic polynomial given by
\begin{equation}
Q(P)=P^3+\frac{9}{8}P^2+\frac{27}{2}P,~~~~P=-\frac{25\epsilon}{b^2}p.
\label{38}
\end{equation} 
The equation $Q(P)=0$ has one real root $P_1=0$ and two complex roots $P_2$ and 
$P_3$ which are complex conjugated. Hence we have the relations: $Q(P)>0$ for $P>0$
and $Q(P)<0$ for $P<0$. The next three cases occur for ${\rm C}_1$ and ${\rm C}_2$ conditions
and an arbitrary parameter $b$ ($b>0$, $b<0$ or $b=0$),

(1) ${\rm C}_1$ condition and $p>0$, or ${\rm C}_2$ condition and $p<0$. In these two cases we have $D(p)<0$ and hence all three roots ${\rm y}_n(p)$ of the polynomial $f({\rm y})$ are real and different. We order the roots as ${\rm y}_1(p)<{\rm y}_2(p)<{\rm y}_3(p)$.

(2) ${\rm C}_1$ condition and $p=0$, or ${\rm C}_2$ condition and $p=0$. In these two cases all three roots ${\rm y}_n(p)$ are real, but two of them are equal because we have $D(p)=0$ for $p=0$. In the case $p=0$ and $\epsilon b<0$ two roots are equal to zero and we order the roots as ${\rm y}_1={\rm y}_2<{\rm y}_3$ where ${\rm y}_1={\rm y}_2=0$ and ${\rm y}_3=-b/5\epsilon>0$.

(3) ${\rm C}_1$ condition and $p<0$, or ${\rm C}_2$ condition and $p>0$. In these two cases we have $D(p)>0$ and hence one root of the polynomial $f({\rm y})$ is real and two roots are complex and they are complex conjugated. We show (section V) that the case (3) leads to unbounded solutions. 

We consider in this section the case (1) and hence $D(p)< 0$. The case (2) arises when the parameter $p$ tends to zero.
In the case (1) the polynomial $f({\rm y})=
({\rm y}_1-{\rm y})({\rm y}_2-{\rm y})({\rm y}_3-{\rm y})$ has three different real roots ${\rm y}_1(p)<{\rm y}_2(p)<{\rm y}_3(p)$. 
First we consider the periodic bounded solution of Eq. (\ref{32}) where ${\rm y}_2\leq{\rm y}\leq{\rm y}_3$.

 One can write Eq. (\ref{32})
as ${\rm d}{\rm y}/{\sqrt{f({\rm y})}}=\pm d{\rm x}$, then 
two substitutions ${\rm y}({\rm x})={\rm y}_3-{\rm s}^2({\rm x})$ and $s({\rm x})=\sqrt{{\rm y}_3-{\rm y}_2}~{\rm z}({\rm x})$ in this equation
lead to the differential equation for the function ${\rm z}({\rm x})$ as
\begin{equation}
\frac{d{\rm z}}{\sqrt{(1-{\rm z}^2)(1-k^2{\rm z}^2)}}=\mp\frac{1}{2}\sqrt{{\rm y}_3-{\rm y}_1}d{\rm x},
\label{39}
\end{equation}
where the parameter $k$ ($0<k<1$) and function ${\rm z}^2({\rm x})$ are 
\begin{equation}
k=\sqrt{\frac{{\rm y}_3-{\rm y}_2}{{\rm y}_3-{\rm y}_1}},~~~~{\rm z}^2({\rm x})=
\frac{{\rm y}_3-{\rm y}({\rm x})}{{\rm y}_3-{\rm y}_2}.
\label{40}
\end{equation}
The integration of Eq. (\ref{39}) with the sign $+$ yields
\begin{equation}
\int_0^{\rm z}\frac{dt}{\sqrt{(1-t^2)(1-k^2t^2)}}=\frac{1}{2}\sqrt{{\rm y}_3-{\rm y}_1}({\rm x}-\eta).
\label{41}
\end{equation}
We use the sign $+$ in this equation because the function ${\rm z}({\rm x})$ is positive when ${\rm x}-\eta$ is small and positive. This equation can be written as
\begin{equation}
F(z;k)={\rm w}({\rm x}-\eta),~~~~{\rm w}=\frac{1}{2}\sqrt{{\rm y}_3-{\rm y}_1},
\label{42}
\end{equation}
where $F(z;k)$ is the Jacobi's form of elliptic integral of the first kind (see Eq. (\ref{4c})).
Thus we have the next solution ${\rm z}({\rm x})={\rm sn}[{\rm w}({\rm x}-\eta),k]$
which yields the function ${\rm y}({\rm x})$ as
\begin{equation}
{\rm y}({\rm x})={\rm y}_3-({\rm y}_3-{\rm y}_2)~{\rm sn}^2[{\rm w}({\rm x}-\eta),k].
\label{43}
\end{equation}
This solution can be written in an equivalent form as
\begin{equation}
{\rm y}({\rm x})={\rm y}_2+({\rm y}_3-{\rm y}_2)~{\rm cn}^2[{\rm w}({\rm x}-\eta),k].
\label{44}
\end{equation}
Here ${\rm sn}(\zeta,k)$ and ${\rm cn}(\zeta,k)$ are the Jacobi's elliptic functions \cite{GR,AS}.
Thus in the case (1) the periodic bounded solution of the generalized NLSE is given by
\begin{equation}
\psi(z,\tau)=[{\rm V}_p+{\rm U}_p~{\rm cn}^2({\rm w}_p\xi,k_p)]\exp[i\Phi(z,\tau)],
\label{45}
\end{equation}
where $\xi=\tau-{\rm v}^{-1}z-\eta$, $\Phi(z,\tau)=\kappa z-\delta\tau+\phi$  and the wave number $\kappa$ is given by Eq. (\ref{16}). The velocity ${\rm v}$ and frequency shift $\delta$ are
\begin{equation}
{\rm v}=\frac{8\epsilon^2}{\sigma(\sigma^2-4\alpha\epsilon)},~~~~\delta=-\frac{\sigma}{4\epsilon}.
\label{46}
\end{equation}
The parameters ${\rm V}_p$, ${\rm U}_p$ and ${\rm w}_p$, $k_p$ of this periodic bounded solution are the functions of integration constant $p$. These parameters have the next explicit form, 
\begin{equation}
{\rm V}_p=\sqrt{-\frac{15\epsilon}{2\gamma}}~{\rm y}_2(p),~~~~{\rm U}_p=\sqrt{-\frac{15\epsilon}{2\gamma}}~({\rm y}_3(p)-{\rm y}_2(p)),
\label{47}
\end{equation}
\begin{equation}
{\rm w}_p=\frac{1}{2}\sqrt{{\rm y}_3(p)-{\rm y}_1(p)},~~~~k_p=\sqrt{\frac{{\rm y}_3
(p)-{\rm y}_2(p)}{{\rm y}_3(p)-{\rm y}_1(p)}},
\label{48}
\end{equation}
where ${\rm y}_n(p)$ are the ordered roots ${\rm y}_1(p)<{\rm y}_2(p)<{\rm y}_3(p)$ of the polynomial $f({\rm y})$. The non-ordered roots ${\rm y}'_n(p)$ of this polynomial for $n=1, 2, 3$ are
\begin{equation}
{\rm y}'_n(p)=2H(p)\cos\left(\frac{\theta}{3}+\frac{2\pi (n-1)}{3}\right),
\label{49}
\end{equation}
where $H(p)=\sqrt{-N(p)/3}$ and the parameter $\theta$ is defined by relation $\cos\theta=-M(p)/2H(p)^3$ with $\sin\theta>0$. Here the functions $N(p)$ and $M(p)$ are given in Eq. (\ref{3c}).
We assume that in this equation $p>0$ for ${\rm C}_1$ condition and $p<0$ for ${\rm C}_2$ condition because the solution in Eq. (\ref{45}) is found for the case (1) with $D(p)<0$.
The period $T_p$ of this periodic bounded solution is
\begin{equation}
T_p=\left(\frac{2}{{\rm w}_p}\right)K(k_p),~~~K(k_p)=\int_0^1\frac{dt}{\sqrt{(1-t^2)(1-k_p^2t^2)}}.
\label{50}
\end{equation}

\begin{figure}
\includegraphics[width=\linewidth,trim={3cm 9cm 3cm 9cm},clip]{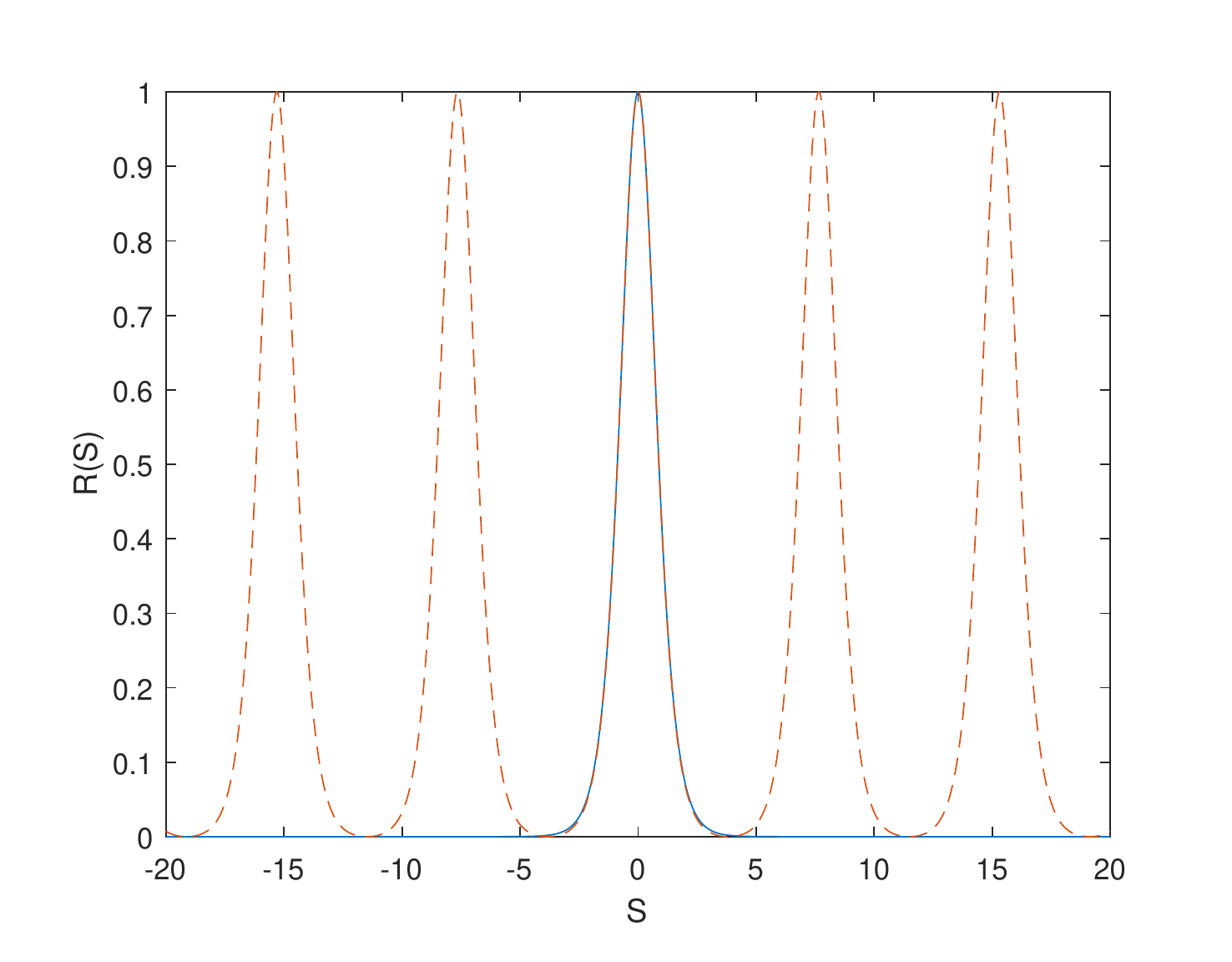}
\caption{Dimensionless function $R(s)$ (dashed line) given by Eq. (\ref{54}) for  parameter $h=10^{-2}$ and soliton solution $R(s)={\rm sech}^2(s)$ (solid line) for parameter $h=0$.
Here $s={\rm w}_0\xi$ is dimensionless variable and $h=\sqrt{P}$ is small parameter.} 
\label{FIG.1.}
\end{figure}
\begin{figure}
\includegraphics[width=\linewidth,trim={3cm 9cm 3cm 9cm},clip]{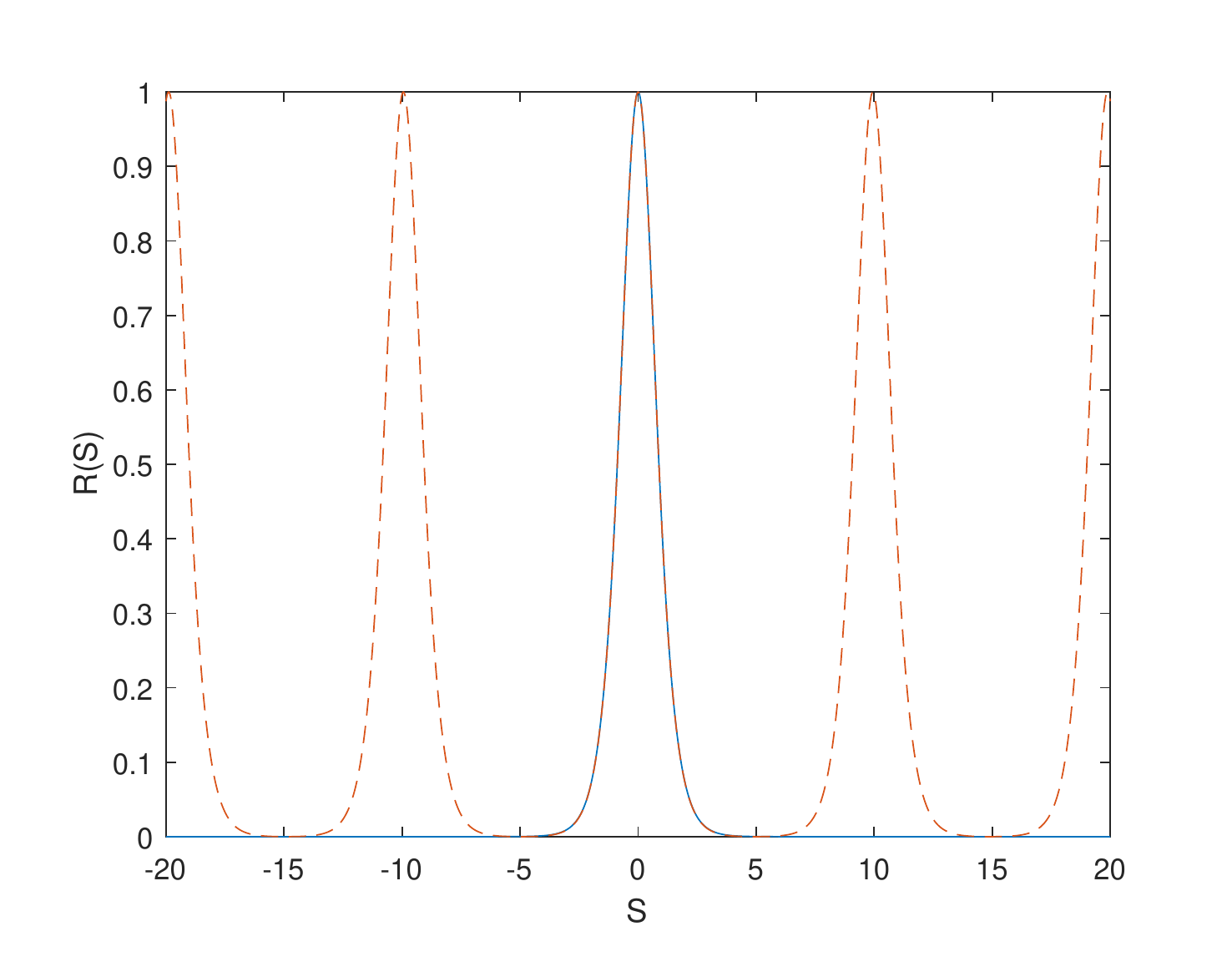}
\caption{Dimensionless function $R(s)$ (dashed line) given by Eq. (\ref{54}) for  parameter $h=10^{-3}$ and soliton solution $R(s)={\rm sech}^2(s)$ (solid line) for parameter $h=0$.
Here $s={\rm w}_0\xi$ is dimensionless variable and $h=\sqrt{P}$ is small parameter.}
\label{FIG.2.}
\end{figure}
In the limiting case when the integration constant $p$ tends to zero
and the conditions $\gamma\epsilon<0$ and $8\alpha\epsilon>3\sigma^2$ are satisfied, Eq. (\ref{48}) yields the next parameters: ${\rm w}_0=\sqrt{-b/20\epsilon}$ and $k_0=1$. Thus the periodic solution in Eq. (\ref{45}) reduces to ${\rm sech}^2$ soliton solution given in Eq. (\ref{20}) when $p=0$, $\gamma\epsilon<0$ and $\epsilon b<0$. 
We refer the solution in Eq. (\ref{45}) as ${\cal P}_{+}$ class of bounded solutions. This class of bounded solutions is obtained for the integration constant $\gamma p\geq0$. The ${\cal P}_{+}$ class of bounded  solutions describes periodic solutions for $\gamma p>0$ and ${\rm sech}^2$ soliton solution for $p=0$, $\gamma\epsilon<0$ and $8\alpha\epsilon>3\sigma^2$.   

In the case (1) with $b=0$ ($8\alpha\epsilon=3\sigma^2$) the polynomial $f({\rm y})$ has the next three real roots: ${\rm y}_1(p)=-\sqrt{-2p/9\epsilon}$, ${\rm y}_2(p)=0$, and ${\rm y}_3(p)=\sqrt{-2p/9\epsilon}$. Hence, in this case the solution in Eq. (\ref{45}) reduces to the periodic bounded solution as
\begin{equation}
\psi(z,\tau)=\sqrt{\frac{5p}{3\gamma}}~{\rm cn}^2({\rm w}_p\xi,k_p)\exp[i(\kappa z-\delta\tau+\phi)],
\label{51}
\end{equation}
where $\xi=\tau-{\rm v}^{-1}z-\eta$, $\delta=-\sigma/4\epsilon$, and the parameters ${\rm v}$ and $\kappa$ are
\begin{equation}
{\rm v}=-\frac{16\epsilon^2}{\sigma^3},~~~~\kappa=p+\frac{3\sigma^4}{256\epsilon^3}.
\label{52}
\end{equation}
In this solution the parameters ${\rm w}_p$ and $k_p$ are given by
\begin{equation}
{\rm w}_p=\left(-\frac{p}{18\epsilon}\right)^{1/4},~~~~k_p=\frac{1}{\sqrt{2}}.
\label{53}
\end{equation}

In the case when $\epsilon b<0$ and $h=\sqrt{P}\ll1$ (see Appendix C) the parameters of solution in  Eq. (\ref{45}) for the second order to small parameter $h$ are given by Eqs. (\ref{10c})-(\ref{12c}). 
We introduce the dimensionless function $R(s)={\rm U}_0^{-1}{\rm u}({\rm x})$ for the ${\cal P}_{+}$ class solutions where $s={\rm w}_0\xi$ is the dimensionless variable and the parameters ${\rm U}_0$ and ${\rm w}_0$
are given by Eq. (\ref{21}).
It follows from Eqs. (\ref{10c})-(\ref{12c}) and Eq. (\ref{45}) that the function $R(s)$ for 
$h\ll1$ has the next form,
\begin{equation}
R(s)={\rm cn}^2(s,k),~~~~k=\left(\frac{9-2\sqrt{3}h+h^2}
{9+2\sqrt{3}h+h^2}\right)^{1/2}.
\label{54}
\end{equation}
This function has the soliton solution $R(s)={\rm sech}^2(s)$ for parameter $h=0$.
We demonstrate the function $R(s)$ (dashed line) in Fig. 1 and Fig. 2  for parameters $h= 10^{-2}$ and $h= 10^{-3}$  respectively and soliton solution $R(s)={\rm sech}^2(s)$ (solid line) for parameter $h=0$. The Fig. 1 and Fig. 2  demonstrate that
the central pulse of the periodic function $R(s)$ and soliton solution $R(s)={\rm sech}^2(s)$ (solid line)  for $h=0$ coincide with a high accuracy.  Hence other pulses of the train given by function $R(s)$ also have ${\rm sech}^2$ soliton shape. 
  
It follows from Eq. (\ref{54}) that in the case (1) provided that $\epsilon b<0$
the asymptotics of the ${\cal P}_{+}$ class solutions, when $h$ tends to zero, is the train of solitons,
\begin{equation}
R(s)\simeq \sum_{n\in{\cal Z}}{\rm sech}^2(s-n{\cal T}),~~~~{\cal T}\simeq \ln(12\sqrt{3}/h),
\label{55}
\end{equation}
where ${\cal Z}=\{0, \pm 1, \pm 2,...\pm\infty\}$.
The period of function $R(s)$ is ${\cal T}={\rm w}_0T_p$ and the asymptotics of function $T_p$ is given by Eq. (\ref{14c}). Hence the asymptotics of period of function $R(s)$ is  ${\cal T}\simeq \ln(12\sqrt{3}/h)$. This equation yields ${\cal T}=7.639$ for parameter $h=10^{-2}$ 
and ${\cal T}=9.942$ for parameter $h=10^{-3}$. We emphasise that these values of ${\cal T}$ and the period of function $R(s)$ in Fig. 1 and Fig. 2 are in a good agreement.
The Fig. 1 and Fig. 2 demonstrate that the train of solitons given by Eq. (\ref{55}) describes the function $R(s)$ with a high accuracy.  

Thus in the case (1) (${\rm C}_1$ condition and $p>0$, or ${\rm C}_2$ condition and $p<0$) provided that $8\alpha\epsilon>3\sigma^2$ the ${\cal P}_{+}$ class of bounded solutions has the asymptotics,
\begin{equation}
\psi(z,\tau)\simeq \sum_{n\in{\cal Z}}{\rm U}_0~{\rm sech}^2[{\rm w}_0(\xi-nT_p)]\exp[i\Phi(z,\tau)],
\label{56}
\end{equation}
where $\xi=\tau-{\rm v}^{-1}z-\eta$, $\Phi(z,\tau)=\kappa z-\delta\tau+\phi,$ and
the wave number $\kappa$ is given by Eq. (\ref{12c}) as
\begin{equation}
~~~\kappa=\kappa_0 -\frac{h^2}
{25\epsilon}\left(\frac{3\sigma^2}{8\epsilon}-\alpha\right)^2.
\label{57}
\end{equation}
Note that this equation can also be written as $\kappa=\kappa_0+p$.
Other parameters in Eq. (\ref{56}) are given by  Eqs. (\ref{21}) and (\ref{22}). 
It follows from Eq. (\ref{14c}) that the asymptotics of 
period of train in Eq. (\ref{56}) is 
\begin{equation}
T_p\simeq \frac{1}{{\rm w}_0}~\ln\frac{12\sqrt{3}}{h},~~~~h=\frac{5\sqrt{-\epsilon p}}{|b|}.
\label{58}
\end{equation}
In the limiting case $h=0$ and hence $p=0$ the train of solitons in Eq. (\ref{56})
reduces to ${\rm sech}^2$ soliton solution given in Eq. (\ref{20}) because the period 
$T_p$ tends to infinity when $h\rightarrow 0$.

These results lead to the next proposition,

{\it Statement 4}. In the case (1) provided that $8\alpha\epsilon>3\sigma^2$ the asymptotics of ${\cal P}_{+}$ class solutions of the generalized NLSE with higher order dispersion is the train of solitons given in Eqs. (\ref{56}) and (\ref{57}). The asymptotics of period $T_p$ is given by Eq. (\ref{58}). In the case $h=0$ ($p=0$) the train of solitons reduces to ${\rm sech}^2$ soliton in Eq. (\ref{20}).

This Statement for the asymptotics of ${\cal P}_{+}$ class of bounded solutions follows directly from Eq. (\ref{45}) and Eqs. (\ref{10c})-(\ref{14c}).

\section{${\cal P}_{-}$ class of unbounded solutions of generalized NLSE}  
 
In this section we describe the ${\cal P}_{-}$ class of unbounded periodic solutions of generalized NLSE which arises in the case (3). Thus this class of unbounded solutions takes place for ${\rm C}_1$ condition and $p<0$, or ${\rm C}_2$ condition and $p>0$. In these both cases the domain of integration constant $p$ for the ${\cal P}_{-}$ class of unbounded solutions is given by relation $\gamma p<0$. In the case (3) we have the relation $D(p)>0$ and hence one root of the polynomial $f({\rm y})$ is real and two roots are complex, they are complex conjugated.
We show below that this ${\cal P}_{-}$ class of unbounded solutions consists of two one-parameter families $\psi_{+}(z,\tau)$ and $\psi_{-}(z,\tau)$ which are expressed in terms of elliptic Jacobi's function. The solutions of Eq. (\ref{32}) depend on 
the roots of the cubic equation $f({\rm y})=0$.  
In the case (3) the roots of this   
cubic equation are 
\begin{equation}
{\rm y}_1(p)=\rho(p)+i\mu(p),~~~~{\rm y}_2(p)=\rho(p)-i\mu(p),
\label{59}
\end{equation}
\begin{equation}
{\rm y}_3(p)=-\frac{b}{15\epsilon}+[\rho_{+}(p)+\rho_{-}(p)],
\label{60}
\end{equation}
where ${\rm y}_1(p)$ and ${\rm y}_2(p)$ are the complex conjugated roots.
In these equations the functions $\rho(p)$, $\mu(p)$ and $\rho_{\pm}(p)$ are 
\begin{equation}
\rho(p)=-\frac{b}{15\epsilon}-\frac{1}{2}[\rho_{+}(p)+\rho_{-}(p)],
\label{61}
\end{equation}
\begin{equation}
\mu(p)=\frac{\sqrt{3}}{2}[\rho_{+}(p)-\rho_{-}(p)],
\label{62}
\end{equation}
\begin{equation}
\rho_{\pm}(p)=\left[-\left(\frac{bp}{135\epsilon^2}+\frac{b^3}{(15\epsilon)^3}\right)\pm\sqrt{D(p)}\right]^{1/3}.
\label{63}
\end{equation}
Let define a new function as ${\rm z}({\rm x})={\rm y}({\rm x})-\rho$
where the parameter $\rho(p)$ is given in Eq. (\ref{61}).
 In this case Eq. (\ref{32}) has the form,
\begin{equation}
\left(\frac{{\rm d}{\rm z}}{{\rm d}{\rm x}}\right)^2=(\nu-{\rm z})(\mu^2+
{\rm z}^2),
\label{64}
\end{equation}
where the parameter $\nu(p)$ is given by
\begin{equation}
\nu(p)=\frac{3}{2}[\rho_{+}(p)+\rho_{-}(p)].
\label{65}
\end{equation}
The solution of Eq. (\ref{64}) for $z\leq \nu$ is 
\begin{equation}
\int_{{\rm z}}^\nu\frac{dt}{\sqrt{(\nu-t)(\mu^2+t^2)}}=\pm({\rm x}-\eta).
\label{66}
\end{equation}
The integral in this equation can be written as
\begin{equation}
\int_{{\rm z}}^\nu\frac{dt}{\sqrt{(\nu-t)(\mu^2+t^2)}}=\frac{1}{\sqrt{\lambda}}F(2\Theta,k),
\label{67}
\end{equation}
where $F(x,k)$ is the elliptic integral of the first kind.
The parameters $\lambda$, $k$ and $\Theta$ in this equation are
\begin{equation}
\lambda=\sqrt{\nu^2+\mu^2},~~~~k=\sqrt{\frac{\lambda+\nu}{2\lambda}},
\label{68}
\end{equation}
\begin{equation}
\Theta={\rm arc}~{\rm tg}~\sqrt{\frac{\nu-{\rm z}}{\lambda}}.
\label{69}
\end{equation}
The integral in Eq. (\ref{67}) can also be written in Jacobi's form by relation
$F(2\Theta,k)=F(G;k)$ where $G$ is
\begin{equation}
G=\sin2\Theta=\frac{2\sqrt{\lambda}~\sqrt{\nu-{\rm z}}}{\lambda+\nu-{\rm z}}.
\label{70}
\end{equation}
Thus Eq. (\ref{66}) has the form,
\begin{equation}
F(G;k)=\pm\sqrt{\lambda}({\rm x}-\eta),~~~~\lambda=\sqrt{\nu^2+\mu^2}.
\label{71}
\end{equation}
It follows from Eq. (\ref{71}) the next equation,
\begin{equation}
\frac{\zeta({\rm x})}{(\lambda+\zeta({\rm x}))^2}=\frac{1}{4\lambda}{\rm sn}^2(\sqrt{\lambda}({\rm x}-\eta),k),
\label{72}
\end{equation}
where $\zeta({\rm x})=\nu-{\rm z}({\rm x})$ is a new function. This equation has two solutions
$\zeta_{+}({\rm x})$ and $\zeta_{-}({\rm x})$
which lead to one-parameter families ${\rm y}_{+}({\rm x})$ and ${\rm y}_{-}({\rm x})$ as
\begin{equation}
{\rm y}_{\pm}({\rm x})=\chi-\lambda\left(\frac{1\pm{\rm cn}(\sqrt{\lambda}({\rm x}-\eta),k)}{1\mp{\rm cn}(\sqrt{\lambda}({\rm x}-\eta),k)}\right),
\label{73}
\end{equation}
where ${\rm y}_{\pm}({\rm x})=\chi-\zeta_{\pm}({\rm x})$ and $\chi=\rho+\nu$. Note that we have the next relation for the elliptic Jacobi's function ${\rm cn}(\sqrt{\lambda}({\rm x}-\eta),k)$ as
\begin{equation}
{\rm cn}(\sqrt{\lambda}({\rm x}-\eta),k)=-{\rm cn}(\sqrt{\lambda}({\rm x}-\eta_0),k),
\label{74}
\end{equation}
where the parameter $\eta_0$ is given by
\begin{equation}
\eta_0=\eta-\frac{2}{\sqrt{\lambda}}K(k),~~~~K(k)=F(\pi/2,k).
\label{75}
\end{equation}
The solutions in Eq. (\ref{73}) can be written as 
${\rm y}_{\pm}({\rm x};\eta)$ then Eq. (\ref{74}) leads to the next relation ${\rm y}_{+}({\rm x};\eta)={\rm y}_{-}({\rm x};\eta_0)$.

Thus in the case (3) (${\rm C}_1$ condition and $p<0$, or ${\rm C}_2$ condition and $p>0$) Eqs. (\ref{2}), (\ref{34}) and (\ref{73}) lead to the next periodic unbounded solutions as
\begin{equation}
\psi_{\pm}(z,\tau)=\left[{\rm X}_p-\Lambda_p~\left(\frac{1\pm{\rm cn}(\bar{{\rm w}}_p\xi,k_p)}{1\mp{\rm cn}(\bar{{\rm w}}_p\xi,k_p)}\right)\right]\exp[i\Phi(z,\tau)],
\label{76}
\end{equation}
where $\xi=\tau-{\rm v}^{-1}z-\eta$ and $\Phi(z,\tau)=\kappa z-\delta\tau+\phi$. The parameters ${\rm v}$, and $\delta$ are given by
Eq. (\ref{46}) and the wave number $\kappa$ is defined in Eq. (\ref{16}). 
The parameters in Eq. (\ref{76}) are defined by the relations, 
\begin{equation}
{\rm X}_p=\sqrt{-\frac{15\epsilon}{2\gamma}}~\chi(p),~~~~\Lambda_p=\sqrt{-\frac{15\epsilon}{2\gamma}}~\lambda(p),
\label{77}
\end{equation}
\begin{equation}
\bar{{\rm w}}_p=\sqrt{\lambda(p)},~~~~k_p=\sqrt{\frac{1}{2}+\frac{\nu(p)}{2\lambda(p)}},
\label{78}
\end{equation}
where the functions $\chi(p)$ and $\lambda(p)$ are given by
\begin{equation}
\chi(p)=-\frac{b}{15\epsilon}+[\rho_{+}(p)+\rho_{-}(p)],
\label{79}
\end{equation}
\begin{equation}
\lambda(p)=\sqrt{3}~[\rho^2_{+}(p)+\rho_{+}(p)\rho_{-}(p)+\rho^2_{-}(p)]^{1/2}.
\label{80}
\end{equation}
The periodic unbounded solutions in Eq. (\ref{76}) can be written as $\psi_{\pm}(z,\tau;\eta)$ then Eq. (\ref{74}) yields the relation $\psi_{+}(z,\tau;\eta)=\psi_{-}(z,\tau;\eta_0)$. We refer the solutions given in Eq. (\ref{76}) for the case (3) as ${\cal P}_{-}$ class of unbounded periodic solutions.

{\it Statement 5}. In the case (3) (${\rm C}_1$ condition and $p<0$, or ${\rm C}_2$ condition and $p>0$) all non-plain wave solutions of the generalized NLSE are the unbounded periodic solutions given by Eq. (\ref{76}). 

This Statement follows directly from the equations of this section. The unbounded solutions are not essential for physical applications. Nevertheless the proposition in this Statement is important because it shows that the non-plain wave bounded solutions of the generalized NLSE exist only in the cases (1) and (2). Thus the ${\cal P}_{+}$ class solutions describes all non-plain wave   bonded solutions of the generalized NLSE given in Eq. (\ref{1}).

\section{Discussions and conclusion}

 A range of new solutions to the nonlinear Schr\"{o}dinger equation with dispersion terms up to fourth order have been found. These include a solitary wave solution given by Eq. (\ref{20}) for integration constant $p=0$, $\gamma\epsilon<0$ and $8\alpha\epsilon>3\sigma^2$. The periodic bounded solutions in Eq. (\ref{45}) arise in the case (1) (${\rm C}_1$ condition and $p>0$, or ${\rm C}_2$ condition and $p<0$). The train of solitary waves given by Eq. (\ref{56}) takes place in the case (1) provided that $8\alpha\epsilon>3\sigma^2$ and $h\ll 1$.  In the case when the integration constant $p$ is zero the train of solitons reduces to the solitary wave solution with ${\rm sech}^2$ shape. It is also shown that no quartic dark soliton solution of the generalized NLSE with higher order dispersion. In the case (3) (${\rm C}_1$ condition and $p<0$, or ${\rm C}_2$ condition and $p>0$) all non-plain wave solutions of the generalized NLSE are the unbounded solutions. These unbounded periodic solutions are given by Eq. (\ref{76}). We emphasize that the solitary wave solution exists only for integration constant $p=0$. Note that this solitary wave solution reduces to the known soliton solution \cite{B2} for $\beta_3=0$. Moreover, it is shown by stability analysis \cite{KRH} and numerical simulsations \cite{MJ} that the temporal shape and the peak power of the soliton are stable when the necessary conditions for existence of ${\rm sech}^2$ soliton are satisfied. This stability analysis is applicable for the arbitrary solitary wave solutions of generalised NLSE with higher order dispersion terms \cite{B5}. It is anticipated that ${\rm sech}^2$ type of stable localized pulses could find practical applications in communications, slow-light devices and ultrafast lasers. The prediction of the periodic trains of solitary pulses may find application in nonlinear optical systems such as lasers and fiber oscillators, while the ${\cal P}_{+}$ class of bounded solutions may find also application in other fields of physics where the nonlinear Schr\"{o}dinger equation including higher order dispersion is a good model of a system.

\section*{Acknowledgements}

The author acknowledge stimulating discussions with John D. Harvey while the support of the Dodd-Walls Centre for Photonic and Quantum Technologies is gratefully acknowledged.

\appendix

\section{Statement 1}

In this Appendix we prove the Statement 1. We write Eqs. (\ref{3}) and (\ref{4}) as
\begin{equation}
r\frac{{\rm d}^3{\rm u}}{{\rm d}{\rm x}^3}-l\frac{{\rm d}{\rm u}}{{\rm d}{\rm x}}=0,~~~~~~~~
\label{1a}
\end{equation}
\begin{equation}
\epsilon\frac{{\rm d}^4{\rm u}}{{\rm d}{\rm x}^4}+\tilde{b}\frac{{\rm d}^2{\rm u}}{{\rm d}{\rm x}^2}-\tilde{c}{\rm u}+\gamma{\rm u}^3=0.~~~~~~~
\label{2a}
\end{equation}
The next notations are used here, 
\begin{equation}
r=\sigma+4\epsilon\delta,~~~~l=-q+2\alpha\delta+3\sigma\delta^2+4\epsilon\delta^3,
\label{3a}
\end{equation}
\begin{equation}
\tilde{b}=-\alpha-3\sigma\delta-6\epsilon\delta^2,~~~~\tilde{c}=\kappa-\alpha\delta^2-\sigma\delta^3-\epsilon\delta^4.
\label{4a}
\end{equation}
First we consider the case when
$\epsilon\neq 0$. Suppose that $r\neq 0$ then 
it follows from Eq. (\ref{1a}) that ${\rm d}^4{\rm u}/{\rm d}{\rm x}^4=(l/r){\rm d}^2{\rm u}/{\rm d}{\rm x}^2$. Hence Eq. (\ref{2a}) reduces to the equation,
\begin{equation}
\left(\tilde{b}+\frac{\epsilon l}{r}\right)\frac{{\rm d}^2{\rm u}}{{\rm d}{\rm x}^2}-\tilde{c}{\rm u}+\gamma{\rm u}^3=0,~~~~~~~
\label{5a}
\end{equation}
Integration of Eq. (\ref{1a}) leads to the next equation,
\begin{equation}
r\frac{{\rm d}^2{\rm u}}{{\rm d}{\rm x}^2}-l{\rm u}=\Omega_0,~~~~~~~~
\label{6a}
\end{equation}
where $\Omega_0$ is the integration constant. Thus the system of Eqs. (\ref{5a})
and (\ref{6a}) yields  
\begin{equation}
\Omega_0\left(\frac{\tilde{b}}{r}+\frac{\epsilon l}{r^2}\right)+\left(\frac{\tilde{b}l}{r}+\frac{\epsilon l^2}{r^2}-\tilde{c}\right){\rm u}+\gamma{\rm u}^3=0,~~~~~~~
\label{7a}
\end{equation}
Hence we have ${\rm u}={\rm constant}$ and ${\rm d}{\rm u}/{\rm d}{\rm x}= 0$ which means that all solutions of the system of Eqs. (\ref{1a})
and (\ref{2a}) are the plain wave solutions when $r\neq 0$. In the case $r=0$ it follows from Eq. (\ref{1a}) the relation $l=0$ for  non-plain wave solutions.
Hence the system of Eqs. (\ref{1a}) and (\ref{2a}) reduces to Eq. (\ref{7}) because in this case $\tilde{b}=b$ and $\tilde{c}=c$. Thus we have obtained
that in the case when $\epsilon\neq 0$ the non-plain wave solutions of the system of Eqs. (\ref{1a}) and (\ref{2a}) exist if and only if 
the next two relations $r=0$ and $l=0$ are satisfied. 
Note that these two relations are equal to the equations given in Eq. (\ref{5}).

In the case when $\epsilon=0$ and $r\neq 0$ Eqs. (\ref{2a}) and (\ref{6a})
lead to equation,
\begin{equation}
\Omega_0\frac{\tilde{b}}{r}+\left(\frac{\tilde{b}l}{r}-\tilde{c}\right){\rm u}+\gamma{\rm u}^3=0.~~~~~~~
\label{8a}
\end{equation}
Hence all solutions of the system of Eqs. (\ref{1a})
and (\ref{2a}) are the plain wave solutions when $\epsilon=0$ and $r\neq 0$. In the case when $r=0$ it follows from Eq. (\ref{1a})
the next relation $l=0$ when the solutions are non-plain waves. Hence we have $\sigma=0$ and $q=2\alpha\delta$ because we assume here that $\epsilon=0$. 
Thus in this case Eq. (\ref{1}) reduces to the ordinary NLSE with second order dispersion term.

\section{Polynomial solution of nonlinear differential equation}

In this Appendix we show that Eq. (\ref{11}) has the polynomial solution.
The substitution of Eq. (\ref{12}) to
Eq. (\ref{11}) leads to the next system of equations for the coefficients $a_n$ and integration constant $\Omega$ as 
\begin{equation}
8a_0a_2-a_1^2+4b\epsilon^{-1}a_0=8\epsilon^{-1}\Omega,
\label{1b}
\end{equation}

\begin{equation}
6a_0a_3+a_1a_2+b\epsilon^{-1}a_1=0,
\label{2b}
\end{equation}

\begin{equation}
9a_1a_3+2a_2^2+2b\epsilon^{-1}a_2-2c\epsilon^{-1}=0,
\label{3b}
\end{equation}

\begin{equation}
5a_2+b\epsilon^{-1}=0,~~~~15a_3^2+2\gamma\epsilon^{-1}=0.
\label{4b}
\end{equation}
This system of equations has the solution,
\begin{equation}
a_0=\frac{2b(\kappa-\kappa_0)}{9\gamma\epsilon},~~~~a_1=\frac{2(\kappa-\kappa_0)}{9\epsilon a_3},
\label{5b}
\end{equation}
\begin{equation}
a_2=-\frac{b}{5\epsilon},~~~~a_3=\pm\sqrt{-\frac{2\gamma}{15\epsilon}},
\label{6b}
\end{equation}
\begin{equation}
\Omega=\frac{3b^2(\kappa-\kappa_0)}{45\gamma\epsilon}+\frac{5(\kappa-\kappa_0)^2}{108\gamma}.
\label{7b}
\end{equation}
Here the parameter $\kappa_0$ is defined as
\begin{equation}
\kappa_0=-\frac{4}{25\epsilon^3}\left(\frac{3\sigma^2}{8}-\alpha\epsilon \right)^2-\frac{\sigma^2}{16\epsilon^3}\left(\frac{3\sigma^2}{16}-\alpha\epsilon \right).
\label{8b}
\end{equation}
Note that the wave number $\kappa$ in Eq. (\ref{8}) is an arbitrary real parameter. Hence the wave number $\kappa$ is the integration constant for traveling solutions given by Eq. (\ref{2}). However we introduce another integration constant as $p=\kappa-\kappa_0$. Thus the 
wave  number $\kappa=p+\kappa_0$ is defined by an arbitrary real parameter $p$ and has the form,
\begin{equation}
\kappa=p-\frac{4}{25\epsilon^3}\left(\frac{3\sigma^2}{8}-\alpha\epsilon \right)^2-\frac{\sigma^2}{16\epsilon^3}\left(\frac{3\sigma^2}{16}-\alpha\epsilon \right).
\label{9b}
\end{equation}
The polynomial solution given in Eq. (\ref{12}) follows from Eq. (\ref{11}) where  $\eta_0=-2\Omega/\epsilon$ and $\eta_1=-c/\epsilon$. Hence the integration constant $\Omega$ and parameter $c$ depend on integration constant $p$ as
\begin{equation}
\Omega=\frac{3b^2p}{45\gamma\epsilon}+\frac{5p^2}{108\gamma},~~~~   c=p-\frac{4b^2}{25\epsilon}.
\label{10b}
\end{equation}

The polynomial solution in Eq. (\ref{12}) leads to Eq. (\ref{17}) where the coefficients $a_n$ and the integration constant $\Omega$ are given in Eqs. (\ref{13}) and (\ref{14}) with $p=\kappa-\kappa_0$.
Note that Eq. (\ref{17}) has the solitary wave solution only in the case when $a_0=0$. Hence
in this case we have $p=0$, and $a_0=a_1=0$ and $\kappa=\kappa_0$. Let 
${\rm u}({\rm x})=\zeta^2({\rm x})$ then Eq. (\ref{17}) has the form,
\begin{equation}
\left(\frac{{\rm d}\zeta}{{\rm d}{\rm x}}\right)^2=\frac{a_2}{4}\zeta^2+\frac{a_3}{4}\zeta^4.
\label{11b}
\end{equation}
In the case $a_2>0$ the localized solution of this equation is given by
\begin{equation}
\zeta({\rm x})=\left(-\frac{a_2}{a_3}\right)^{1/2}{\rm sech}\left(\frac{\sqrt{a_2}}{2}({\rm x}-\eta)\right).
\label{12b}
\end{equation}
In the case $a_2<0$ we have the unbounded solution as
\begin{equation}
\zeta({\rm x})=\left(-\frac{a_2}{a_3}\right)^{1/2}{\rm sec}\left(\frac{\sqrt{-a_2}}{2}({\rm x}-\eta)\right).
\label{13b}
\end{equation}

\section{Asymptotics of ${\cal P}_{+}$ class of bounded solutions}

In this Appendix we present the supplementary material for ${\cal P}_{+}$ class of bounded solutions of generalized NLSE. 
Note that equation $f({\rm y})=0$ has the form,
\begin{equation}
{\rm y}^3-c_2{\rm y}^2-c_1{\rm y}-c_0=0,
\label{1c}
\end{equation}
where the coefficients $c_n$ are given in Eq. (\ref{33}).
It is known that the character of roots in this equation depends on the sign of expression,
\begin{equation}
D(p)=\frac{1}{4}M^2(p)+\frac{1}{27}N^3(p),
\label{2c}
\end{equation}
where $N(p)$ and $M(p)$ are
\begin{equation}
N(p)=-c_1-\frac{c_2^2}{3},~~~~M(p)=-c_0-\frac{c_1c_2}{3}-\frac{2c_2^3}{27}.
\label{3c}
\end{equation}
There are three different cases: 

(1)  Three roots are real and different for $D(p)<0$. 

(2)  Three roots are real, but two of them are equal for $D(p)=0$. 

(3)  One root is real and other two roots are complex and complex conjugated for $D(p)>0$. 

We consider in this Appendix the case (1) and hence $D(p)< 0$. In this case (1) the generalised NLSE leads to the ${\cal P}_{+}$ class of bounded solutions which is expressed in terms of Jacobi's elliptic functions.
We use the Jacobi's form of the elliptic function given by   
\begin{equation}
F({\rm z};k)=\int_0^{\rm z}\frac{dt}{\sqrt{(1-t^2)(1-k^2t^2)}}={\rm arc}~{\rm sn}({\rm z},k).
\label{4c}
\end{equation}
In addition we have the relation $F({\rm z};k)=F(\theta,k)$ where ${\rm z}=\sin \theta$ and 
$F(\theta,k)$ is the elliptic function of the first kind defined by
\begin{equation}
F(\theta,k)=\int_0^{\theta}\frac{d\phi}{\sqrt{1-k^2\sin^2\phi}}.
\label{5c}
\end{equation}
Note that the Jacobi's elliptic functions have different notations in Refs. \cite{GR,AS}.

Equation (\ref{1c}) has the simple solutions for enough small integration constant $p$. The parameter $p$ and Eq. (\ref{1c}) can be presented in the dimensionless form.
We define the dimensionless parameter $P$ and dimensionless variable ${\rm Y}$ as
\begin{equation}
P=-\frac{25\epsilon}{b^2}p,~~~~{\rm Y}=-\frac{15\epsilon}{b}{\rm y},
\label{6c}
\end{equation}
where it is assumed that $b\neq 0$.
Thus Eq. (\ref{1c}) has the next dimensionless form,
\begin{equation}
{\rm Y}^3-3{\rm Y}^2-2P{\rm Y}+4P=0.
\label{7c}
\end{equation}
It follows from Eqs. (\ref{37}) and (\ref{38}) that three roots of Eq. (\ref{1c}) are real and different for the condition ${\rm C}_1$ with $p>0$ and for the condition ${\rm C}_2$ with $p<0$.
Hence the dimensionless parameter $P$ is positive for these two cases. 
We consider below the case when three roots of Eq. (\ref{1c}) are real and different, and the parameter $h=\sqrt{P}$ is small. The perturbation theory yields the roots of Eq. (\ref{7c}) in the second order to small parameter $h$ as
\begin{equation}
{\rm Y}_1=-\frac{2h}{\sqrt{3}}-\frac{h^2}{9},~~~
{\rm Y}_2=\frac{2h}{\sqrt{3}}-\frac{h^2}{9},~~~
{\rm Y}_3=3+\frac{2h^2}{9}.
\label{8c}
\end{equation}
The roots of  Eqs. (\ref{1c}) and (\ref{7c}) are connected as
\begin{equation}
{\rm y}_n(p)=-\frac{b}{15\epsilon}{\rm Y}_n(p),~~~~h=\frac{5\sqrt{-\epsilon p}}{|b|}.
\label{9c}
\end{equation}
Thus in the second order to small parameter $h$, the solution in Eq. (\ref{45}) has the next parameters, 
\begin{equation}
{\rm V}_p={\rm U}_0\left(\frac{2h}{3\sqrt{3}}-\frac{h^2}{27} \right),~~~{\rm U}_p={\rm U}_0\left(1-\frac{2h}{3\sqrt{3}}+\frac{h^2}{9}\right),
\label{10c}
\end{equation}
\begin{equation}
{\rm w}_p={\rm w}_0\left(1+\frac{2h}{3\sqrt{3}}+\frac{h^2}{9}\right)^{1/2},~~~{\rm w}_0=\frac{1}{4}
\sqrt{\frac{4\alpha}{5\epsilon}-\frac{3\sigma^2}{10\epsilon^2}},
\label{11c}
\end{equation}
\begin{equation}
k_p=\left(\frac{9-2\sqrt{3}h+h^2}
{9+2\sqrt{3}h+h^2}\right)^{1/2},~\kappa=\kappa_0-\frac{h^2}
{25\epsilon}\left(\frac{3\sigma^2}{8\epsilon}-\alpha\right)^2,
\label{12c}
\end{equation}
where the amplitude ${\rm U}_0$ is given by  Eq. (\ref{21}). Note that we have found these parameters in the case when the next condition $8\alpha\epsilon>3\sigma^2$ is satisfied. It is shown in section IV that these parameters in Eq. (\ref{45}) yield 
the train of solitons for $h\ll 1$. 

Let consider the case $1-k_p^2\ll 1$ then we have the next asymptotic relation \cite{GR},
\begin{equation}
K(k_p)\simeq \ln\frac{4}{\sqrt{1-k_p^2}}\simeq \ln\frac{6}
{3^{1/4}h^{1/2}}, 
\label{13c}
\end{equation}
where $h\ll 1$.
The period of the ${\cal P}_{+}$ class of bounded solutions is $T_p=(2/{\rm w}_p)K(k_p)$. Hence 
the asymptotic equation for the period $T_p$ is
\begin{equation}
T_p\simeq \frac{2}{{\rm w}_0}~\ln\frac{6}
{3^{1/4}h^{1/2}}=\frac{1}{{\rm w}_0}~\ln\frac{12\sqrt{3}}{h}. 
\label{14c}
\end{equation}
In the case when $h\rightarrow 0$ the period $T_p$ tends to infinity and the ${\cal P}_{+}$ class of bounded solutions reduces to ${\rm sech}^2$ soliton solution given in Eq. (\ref{20}).


\begin{thebibliography}{99}
\bibitem{NS}  The necessary condition $8\alpha\epsilon>3\sigma^2$ for the existence of ${\rm sech}^2$ solitary wave shows that the solitary wave solution of the generalized NLSE does not exist when the second order dispersion is zero.
\end{thebibliography}
\end{document}